# A scaling theory for the size distribution of emitted dust aerosols suggests climate models underestimate the size of the global dust cycle

Jasper F. Kok[1]

[1]Advanced Study Program, National Center for Atmospheric Research, Boulder, Colorado, USA

**ABSTRACT**

Mineral dust aerosols impact Earth's radiation budget through interactions with clouds, ecosystems, and radiation, which constitutes a substantial uncertainty in understanding past and predicting future climate changes. One of the causes of this large uncertainty is that the size distribution of emitted dust aerosols is poorly understood. The present study shows that regional and global circulation models (GCMs) overestimate the emitted fraction of clay aerosols (< 2 μm diameter) by a factor of ~2 – 8 relative to measurements. This discrepancy is resolved by deriving a simple theoretical expression of the emitted dust size distribution that is in excellent agreement with measurements. This expression is based on the physics of the scale-invariant fragmentation of brittle materials, which is shown to be applicable to dust emission. Because clay aerosols produce a strong radiative cooling, the overestimation of the clay fraction causes GCMs to also overestimate the radiative cooling of a given quantity of emitted dust. On local and regional scales, this affects the magnitude and possibly the sign of the dust radiative forcing, with implications for numerical weather forecasting and regional climate predictions in dusty regions. On a global scale, the dust cycle in most GCMs is tuned to match radiative measurements, such that the overestimation of the radiative cooling of a given quantity of emitted dust has likely caused GCMs to underestimate the global dust emission rate. This implies that the deposition flux of dust and its fertilizing effects on ecosystems may be substantially larger than thought.

**Introduction**

Mineral dust aerosols eroded from arid soils impact weather and climate by scattering and absorbing radiation (1-4) and by modifying cloud properties (1, 5). Deposition of dust aerosols also partially controls the productivity and carbon sequestration of ocean ecosystems by providing limiting micronutrients such as iron, which affects atmospheric concentrations of greenhouse gases (6). The total impact of dust aerosols on Earth's radiative budget constitutes an important uncertainty in understanding past and predicting future climate changes (1, 6-8). In addition, dust aerosols adversely affect human health (9) and could suppress hurricane activity (10).

All these processes depend on the size of the atmospheric dust aerosols (2-5), which also determines their lifetime (3). But current treatments of the particle size distribution (PSD) of emitted dust aerosols in global circulation models (GCMs) are based on empirical relations with limited or no physical basis (3, 11-13). This use of empirical relations is necessary both because of the scarcity of measurements (14, 15), and because the understanding of the physical processes that determine the emitted dust PSD is very limited (16, 17). As a consequence, the fraction of emitted dust aerosols in the clay size range (< 2 μm diameter), which both interact most efficiently with shortwave (solar) radiation and have the longest lifetime, differs by up to a factor of 4 between GCMs (3, 11-13, 18-20). The availability of an accurate expression for the emitted dust PSD could thus reduce the uncertainty on GCM estimates of dust climate forcing. The present study derives such an expression from the analogy between the fragmentation of soil dust aggregates and the much better understood fragmentation of brittle materials such as glass (21). The resulting theoretical expression for the emitted dust PSD is in excellent agreement with measurements. In contrast, GCMs overestimate the emitted fraction of clay aerosols by a factor of ~2 – 8, with implications for simulations of the spatial distribution, radiative forcing, and global emission rate of dust aerosols.

**The physics of dust emission**

Dust aerosols that undergo long-range transport predominantly have diameters smaller than 20 μm (22), and are denoted here as PM20 dust. The cohesive forces on such small particles in soils are generally much larger than aerodynamic forces (22), thereby preventing PM20 dust from being lifted directly by wind (15, 22). Moreover, these strong cohesive forces cause PM20 dust to rarely occur loosely in soils because they easily attach to other particles, thereby forming dust aggregates of larger sizes (17, 23).

Instead of being lifted directly by wind, PM20 dust is generally emitted by an intermediary process called saltation (22, 24). In saltation, larger sand-sized particles (~70 – 500 μm (22)), which are more easily lifted by wind because their cohesive forces are small compared to aerodynamic forces (22), move in ballistic trajectories (24). Upon impact on the soil bed, these saltating particles can eject dust particles from dust aggregates in the soil (22), a process known as sandblasting (22, 23). In addition, some saltating particles are sand-sized aggregates of dust particles, which can fragment and emit dust aerosols upon striking the surface (17).

Although the processes leading to dust emission are qualitatively understood, a detailed quantitative understanding is hindered by the large, highly variable, and poorly understood cohesive forces on PM20 dust in soils (16, 17, 22). As a consequence, previous physically-based theories of dust emission that account for cohesive forces (16, 17) have large uncertainties in their input parameters and can differ greatly from measurements (17). I therefore take a different approach here and utilize the closest analog problem to dust emission that is quantitatively understood: the fragmentation of brittle materials (21, 25).

## The fragmentation of brittle materials

When a brittle material such as glass or gypsum receives a large input of energy, for example by being dropped on a hard surface (26), the resulting fragment PSD can be classified into three regimes (21, 27):

1. The elastic regime: for low energies, the brittle object remains intact and no fragments are produced.
2. The damage regime: for larger energies, the object breaks into 2 or more fragments, but the size of the largest fragment is comparable to the original size of the object.
3. The fragmentation regime: for even larger energies, the brittle object fragments into a wide range of particle sizes, for which the size of the largest fragment is small compared to that of the original object.

Measurements show that the PSD in the fragmentation regime follows a power law (21, 26, 28, 29), and is thus 'scale invariant' (30) (Fig. 1). Specifically, measurements and theory show that the PSD arising from the fragmentation of spherical, brittle objects is well-described by (21, 28)

$$\frac{dN_f}{d \ln D_f} \propto D_f^{-2}, \qquad (x_0 < D_f < \lambda) \qquad (1)$$

where $N_f$ is the number of fragments with size $D_f$. This power law is valid for fragments larger than the indivisible constituent size $x_0$ and smaller than the side crack propagation length $\lambda$, the physical significance of which are discussed below.

The reason for the occurrence of scale invariance in brittle material fragmentation is rooted in the mechanism by which fragments are created: through the propagation and merger of cracks. When a brittle object is stressed, for example, by a physical impact, such cracks can propagate along imperfections that locally weaken the material, such as microscopic cracks (21, 29, 31). If the stress is large enough, the material will fail at the tips of the microscopic cracks, causing a larger crack to propagate perpendicular to the stress direction, much like ripping a tear in a piece of fabric. As a result, the length of the crack increases, which in turn increases the stress and thus the crack propagation speed (31). This positive feedback continues until a critical stress is reached at which the main crack generates a side crack, which propagates perpendicular to the direction of the main crack (21). The formation of a side crack temporarily decreases the stress on the main crack.

However, the stress increases again with the crack length as the main crack propagates further, until the stress becomes sufficient to generate another side crack. This repetitive process causes the main crack to produce $N_b$ side cracks at quasi-regular intervals $l_b$ (21). These side cracks propagate as well and, because the stress between side cracks increases with decreasing distance between them (21, 31), are attracted to their nearest neighbor. For a 2-dimensional object, the resulting merger of side cracks produces $N_b/2$ fragments of typical size $l_b$ (21). Yet another crack propagates from the merger of two side cracks, and the resulting $N_b/2$ side cracks of the next generation are also attracted to their nearest neighbor, and form $N_b/4$ fragments of typical size $2l_b$ when they join. This process continues until no more side cracks remain, resulting in a scale-invariant PSD that, for a 3-dimensional spherical object, is approximately given by Eq. (1) (21, 29). Analogous processes produce scale invariance in a large variety of natural systems, including avalanches (30), and the fragmentation of rocks (29) and atomic nuclei (32).

The scale invariance produced by the merger of cracks can however only hold for an intermediate regime $x_0 < D_f < \lambda$, because the number of particles per infinitesimal bin diverges for $D_f \to 0$ and, similarly, the mass contained in each infinitesimal bin diverges for $D_f \to \infty$. Indeed, physically Eq. 1 cannot hold for very small particle sizes, since the 'indivisible' constituents of the material (molecules, crystal cells, or individual dust particles in the case of dust aggregates - see below) have a finite length scale $x_0$ that prevents the creation of fragments of smaller sizes.

Whereas indivisible constituents prevent scale invariance at very small fragment sizes, the limited length $\lambda$ to which side cracks propagate prevents scale variance at large fragment sizes (21, 29). Experiments show that, for fragile brittle materials in which the dissipation of the side crack propagation is not limiting the creation of fragments (29), the resulting large-size cutoff can be approximately described by the product of the power law (Eq. 1) with an exponential function in terms of the fragment volume (26, 29):

$$\frac{dN_f}{d \ln D_f} \propto D_f^{-2} \exp\left[-\left(\frac{D_f}{\lambda}\right)^3\right]. \qquad (D_f > x_0) \qquad (2)$$

Although the side crack propagation length $\lambda$ is generally on the order of 10% of the size of the original object (26, 33), its exact value depends on the density of main cracks in the material. If a large number of main cracks are nucleated for a given sample size, then the side cracks cannot extend far before encountering another main crack and its side cracks, causing $\lambda$ to be small. A larger input of energy nucleates more main cracks, and therefore results in the large-size cutoff moving to smaller particle sizes (21, 27, 29).

## Theoretical model of dust emission

A brittle material is defined as "a material that is, more or less, linearly elastic up to a breaking strain where elasticity vanishes. […] At breaking the (local) material failure is complete, rapid, and (almost always) irreversible" (21). When stressed, dry soil aggregates usually fail as brittle materials (34-36). But, whereas the propagation of cracks in conventional brittle materials like glass is dependent on material imperfections that locally weaken the material, the weakest points in dust aggregates are the surfaces of the constituent dust particle. As a consequence, the propagation of cracks in dust aggregates will proceed along the surfaces of the constituent dust particles.

When a saltating particle impacts a dust aggregate, the resulting PSD of fragments will fall into either the elastic, damage, or fragmentation regimes (defined above), depending on the impact energy and the dust aggregate cohesiveness (27). In order to predict the PSD of emitted dust aerosols, we therefore need to determine which regime is the dominant contributor to dust emission. Although the damage regime produces fragments, the fragmentation regime produces a much larger number of fragments per collision (27).

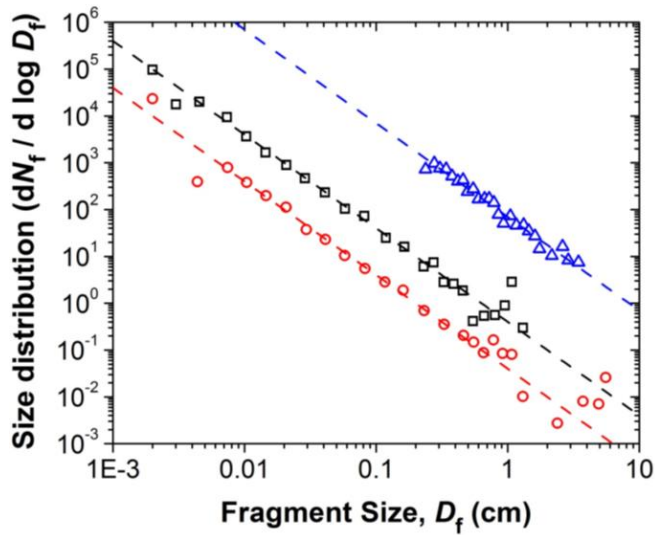

Fig. 1. Size distribution resulting from the fragmentation of a gypsum ball (blue triangles, taken from ref. (26)) and two glass spheres (taken from ref. (28)) of 2.4 cm (red circles) and 7.4 cm (black squares). The dashed lines denote the corresponding power laws with an exponent of -2 (see Eq. 1).

Moreover, each impact in the damage regime nucleates cracks, thereby weakening the dust aggregate and increasing the likelihood that a subsequent impact will result in the fragmentation of the aggregate (31). For the above reasons, the present study assumes that, for most conditions, dust emission is predominantly due to fragmenting impacts. This assumption is further justified by the occurrence of scale invariance in the emission of dust aerosols (see below), which is unique to the fragmentation regime (27).

**Fragmentation of dust aggregates.** Although the physics of dust aggregate fragmentation is expected to be similar to the fragmentation of conventional brittle materials like glass, there is one important difference: the size of the indivisible constituents. In conventional brittle materials, these indivisible constituents are usually individual molecules or crystal units with sizes on the order of $10^{-10} - 10^{-9}$ m, such that the small-size deviation from the power law of Eq. (1) due to the presence of indivisible constituents is only rarely observed (29). However, the indivisible constituents of dust aggregates are the discrete dust particles with much larger typical sizes of $10^{-7} - 10^{-5}$ m (16, 17, 23). As a consequence, we can expect the small-size deviation from the power law to occur at a typical fragment size of ~1 μm.

Many dust aerosols are aggregates themselves (17, 23, 37), such that an emitted dust aerosol consists of one or more soil particles of equal or smaller size. The production of dust aerosols with size $D_d$ is thus proportional to the volume fraction of soil particles with size $D_s \leq D_d$ that contribute to the formation of these aerosols. That is, if the fraction of soil particles with size $D_s \leq D_d$ is doubled, then the production of aerosols with size $D_d$ will be doubled as well, provided that the shape of the soil PSD remains constant. I therefore neglect any influence of changes in the shape of the soil PSD and assume that the production of aerosols of size $D_d$ is proportional to the volume fraction of soil particles with size $D_s \leq D_d$,

$$\frac{dN_d}{d \ln D_d} \propto \int_0^{D_d} P_s(D_s) dD_s \quad (3)$$

where $N_d$ is the normalized number of emitted dust aerosols with size $D_d$, and $P_s$ is the fully-dispersed PSD of PM20 soil particles. The distribution of fully-disaggregated soil particles is usually described as a log-normal distribution (25):

$$P_s(D_s) = \frac{1}{\sqrt{2\pi} \ln(\sigma_s)} \exp\left\{-\frac{\ln^2(D_s / \overline{D_s})}{2 \ln^2(\sigma_s)}\right\}, \quad (4)$$

where $\sigma_s$ is the geometric standard deviation, and $\overline{D_s}$ is the median diameter by volume (2). The number and volume size distributions of emitted dust aerosols are then obtained by combining Eqs. 2 - 4,

$$\frac{dN_d}{d \ln D_d} = \frac{1}{c_N D_d^2}\left[1 + \text{erf}\left(\frac{\ln(D_d / \overline{D_s})}{\sqrt{2} \ln \sigma_s}\right)\right]\exp\left[-\left(\frac{D_d}{\lambda}\right)^3\right], \quad (5)$$

$$\frac{dV_d}{d \ln D_d} = \frac{D_d}{c_V}\left[1 + \text{erf}\left(\frac{\ln(D_d / \overline{D_s})}{\sqrt{2} \ln \sigma_s}\right)\right]\exp\left[-\left(\frac{D_d}{\lambda}\right)^3\right], \quad (6)$$

where $V_d$ is the normalized volume of dust aerosols with size $D_d$, $c_N$ and $c_V$ are normalization constants, and erf is the error function (see pp. 423 of ref. (2)).

Note that Eqs. (5, 6) are only applicable when dust emission is predominantly due to the fragmentation of soil aggregates. Eqs. (5, 6) are for example not valid for (i) aerodynamically lifted dust (22), (ii) dust emitted mainly by impacts in the damage regime, which could occur for very cohesive soils, and (iii) dust with diameters larger than ~20 μm, which are more likely to occur as loose particles in the soil (17, 23), such that their emission is not always due to fragmenting impacts.

**The fully-dispersed PSD of arid soils.** Eqs. (5, 6) depend on the log-normal distribution parameters of fully-dispersed soil particles with sizes extending from the submicron range up to 20 μm. Although measurements of soil PSDs extending into the submicron range have been published (38, 39), there are not nearly enough measurements to parameterize a spatially-varying soil PSD in GCMs. To circumvent this problem, we must define a 'typical' PM20 arid soil PSD.

Parameterizing a typical PM20 arid soil PSD is difficult, however, because there are few published arid soil PSDs that extend into the submicron range. A notable exception is the study of d'Almeida and Schütz (40), who reported such PSDs for six arid soils from across the Sahara (Table 1). These authors used wet sieving and optical and electron microscopy to determine the soil PSD from ~0.01 – 1000 μm. Although the water in which d'Almeida and Schütz suspended their samples dispersed the soil to some degree, they did not fully disperse their samples, for example, by ultrasonic shaking or using a dispersing agent (38, 39). However, measurements suggest that the difference between the fully-dispersed soil PSD and that obtained from suspension in water could be limited, especially in the clay size range (41).

Table 1: Values of the log-normal soil PSD parameters $\overline{D_s}$ and $\sigma_s$, obtained using least-squares fitting of Eq. 4 to the PSDs of 8 arid soils with a range of textures and geographical origins (40, 42)

| Study | Soil number | Soil texture | Geographical location | Best fit $\overline{D_s}$ (μm) | Best fit $\sigma_s$ |
|---|---|---|---|---|---|
| Ref. (40) | 1 | Loam | Mali | 2.6 | 2.9 |
| Ref. (40) | 2 | Silt | Senegal | 1.6 | 3.4 |
| Ref. (40) | 3 | Sand | Mali | 1.7 | 2.8 |
| Ref. (40) | 4 | Loamy sand | Algeria | 7.2 | 3.7 |
| Ref. (40) | 5 | Sand | Niger | 2.1 | 2.9 |
| Ref. (40) | 6 | Sandy loam | Sudan | 4.9 | 2.7 |
| Ref. (42) | 00-U36 | Sand | Utah | 3.0 | 2.8 |
| Ref. (42) | 00-U37 | Loam | Utah | 3.8 | 2.8 |
| Average and standard deviation | | | | 3.4 ± 1.9 | 3.0 ± 0.4 |

I thus parameterize a typical PM20 arid soil PSD from the measurements of d'Almeida and Schütz (40), supplemented by laser diffraction measurements of actively eroding soils in Utah (42). A least-squares fitting technique was used to determine the most likely values of the log-normal parameters $\overline{D_s}$ and $\sigma_s$ for each individual soil (Table 1). The values of these parameters are surprisingly uniform, despite the wide range of soil textures and geographical locations represented. The average and standard deviation of the log-normal parameters of the 8 soil PSDs are $\overline{D_s}$ = 3.4 ± 1.9 μm and $\sigma_s$ = 3.0 ± 0.4, and the log-normal PSD with these average parameters appears to be a reasonable approximation to the 8 available arid soil PSDs (see SI Fig. 1). However, the accuracy of the average values of $\overline{D_s}$ and $\sigma_s$ estimated in Table 1 remains uncertain because 6 of the 8 soil PSDs are not fully dispersed (40), and because of the small number of soils represented.

**Measurements of the vertical dust flux.** Testing the validity of Eqs. (5, 6) requires measurements of the emitted dust PSD. Although measurements of the PSD of atmospheric dust aerosols are relatively abundant in the literature (37, 40, 43, 44), these measurements are not representative of the emitted dust PSD because they inherently include aerosol removal due to deposition. We instead require measurements of the size-resolved vertical dust flux produced by arid soils, which is a direct measure of the PSD of the emitted dust aerosols (45). A list of published size-resolved vertical dust flux measurements is given in SI Table 1.

To allow size-resolved vertical dust flux measurements from the 6 distinct soils investigated in the literature (14, 15, 45, 46) to be collectively compared against Eq. 5, the measurements were processed as follows. Because measurements follow the power law of Eq. 1 in the range of 2 – 10 μm (see inset of Fig. 2 and ref. (15, 45)), each set of measurements in that size range for a given soil and a given wind speed were fit to this power law. Measurements at all aerosol sizes for this given soil and wind speed were then normalized by the proportionality constant in the fitted power law to eliminate the strong dependence of the dust flux on the wind speed. For a given soil, this procedure put measurements at different wind speeds on an equal footing, except for the dependence of the shape of the dust PSD on the wind speed, which measurements suggest to be small (15, 44, 46). The normalized measurements at the various wind speeds for a given soil were then averaged for each particle size to reduce measurement noise and obtain the standard error. Since ref. (45) obtained only one reliable measurement per particle size, the standard error on these single measurements was estimated from the similar measurements of refs. (15, 46).

**Results**

The theoretical emitted dust PSD (Eqs. 5, 6) depends on the parameters $\overline{D_s}$, $\sigma_s$, and $\lambda$. The soil parameters $\overline{D_s}$ = 3.4 μm and $\sigma_s$ = 3.0 are taken from Table 1, and the side crack propagation length $\lambda$ = 12 ± 1 μm is obtained from a least-squares fit to the measurements in Fig. 2. This latter result is in agreement both with the expected value of ~10 % (26, 33) of the typical dust aggregate size of ~20 – 300 μm (17), and with the occurrence of scale invariance in dust emission up to a particle diameter of ~10 μm (see inset of Fig. 2 and discussion below). These values of $\overline{D_s}$, $\sigma_s$, and $\lambda$ yield $c_N$ = 0.9539 μm and $c_V$ = 12.62 μm for the normalization constants in Eqs. (5, 6).

**Comparison of theory with measurements.** Fig. 2 shows that the PSD of emitted dust aerosols is indeed scale invariant in the range of 2 – 10 μm, since it closely follows the predicted power law of Eq. 1. Moreover, the PSD is reduced relative to the power law for small particle sizes (< 2 μm), as predicted from the dust aggregate indivisible constituent size of ~1 μm. The emitted dust PSD is also reduced relative to the power law for larger particle sizes (> 10 μm), as predicted from the limited propagation length of side cracks. Measurements of the emitted dust PSD thus provide strong qualitative support for the dust emission theory presented above.

In addition to this qualitative agreement, Fig. 2 shows excellent quantitative agreement between theory and measurements. Do note that the theory is poorly constrained in the submicron range because of sparse measurements (14).

The final important result evident from Fig. 2 is the small amount of scatter between the dust flux data sets, even though these data were obtained for widely varying wind and soil conditions (SI Table 1). This similarity suggests that changes in the wind and soil conditions have only a limited effect on the emitted dust PSD, as also suggested by the insensitivity of dust aerosol PSDs to changes in wind speed and source region (37, 43, 44). Although more research is needed to fully verify this hypothesis, the apparent insensitivity of the emitted dust PSD to specific soil and wind conditions is highly fortuitous for regional and global dust modeling (44).

**Comparison of theory with empirical model relations.** Fig. 3 compares the theoretical expression of Eq. 6 to empirical relations used in regional models and GCMs. The theory is clearly an improvement over empirical model relations, since it provides much better agreement with measurements. In fact, both theory and measurements show that empirical model relations substantially overestimate the mass fraction of emitted clay aerosols ($D_d$ < 2 μm), whereas they underestimate the fraction of emitted large silt aerosols ($D_d$ > ~5 μm). The greater fraction of large silt aerosols found by measurements and predicted by theory is consistent with the underestimation of the long-range transport of large silt aerosols by GCMs (11, 12, 47).

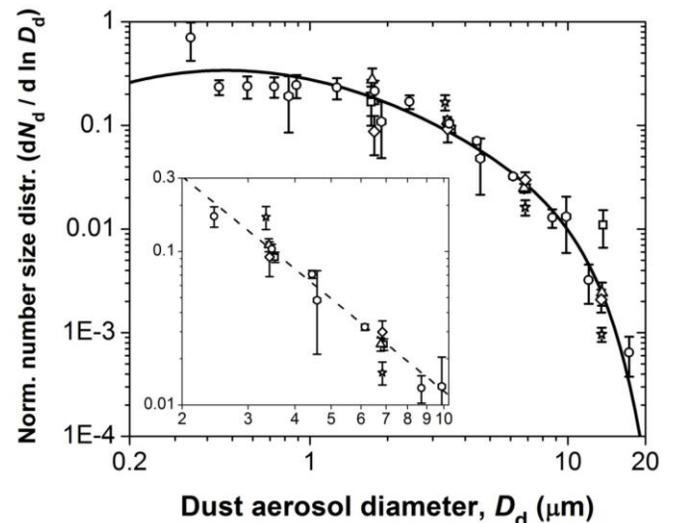

Fig. 2. Measurements with standard error of the normalized number size distribution of emitted dust aerosols [circles (14), squares (15), hexagons (45), triangles (soil 1 in ref. (46)), stars (soil 2 in ref. (46)), and diamonds (soil 3 in ref. (46))]. The inset shows measurements in the size range of 2 – 10 μm only. The solid line denotes the theoretical prediction of Eq. 5, and the dashed line denotes the power law of Eq. 1 observed for the fragmentation of brittle materials (see Fig. 1).

With the exception of ref. (48), GCMs assume emitted clay mass fractions ranging from ~10 % (11, 12, 20, 49) to ~35 % (13, 18), whereas measurements and theory indicate an emitted clay fraction of 4.4 ± 1.0 % for the average values and standard errors of $\overline{D_s}$, $\sigma_s$, and $\lambda$ listed above and in Table 1. Likewise, measured PSDs of dust aerosols indicate an atmospheric clay fraction of ~10 – 20 %, whereas GCMs predict an atmospheric clay fraction of ~20 – 60 % (SI Table 2). Further evidence that GCMs overestimate the emitted clay fraction was reported by Cakmur et al. (49), who found that optimal agreement of a GCM with measurements (e.g., dust aerosol optical depth, deposition, and PSD) requires a smaller clay fraction than normally used.

**Implications for regional and global dust modeling**

Since the lifetime and radiative properties of clay and silt aerosols differ substantially, the overestimation of the clay fraction has implications for regional and global dust modeling. Whereas silt aerosols ($D_d$ > 2 μm) have lifetimes up to a few days and can produce either a positive or a negative forcing by absorbing and scattering both shortwave and longwave radiation, clay aerosols ($D_d$ < 2 μm) have lifetimes on the order of a week and produce a strong negative radiative forcing by efficiently scattering shortwave radiation (3, 4, 18-20, 48). The overestimation of the clay fraction thus causes model errors in both the spatial distribution of dust and in the balance between dust radiative cooling and heating. On local and regional scales in dusty regions, this affects the magnitude of the dust radiative forcing and, depending on other factors such as the surface albedo and the height of the dust (4), can even change the sign of the radiative forcing (50). These effects have implications for numerical weather forecasting as well as for regional climate predictions in dusty regions, especially if dust emissions change substantially in future climates, as hypothesized (7).

To investigate the effect of the clay fraction overestimation on simulations of the dust cycle on global scales, I obtain the mass-normalized radiative forcing for individual particle size bins from a recent GCM study (19) (see SI Fig. 2). (Note that refs. (18, 20, 48) also report the radiative forcing of individual particle size bins, but these studies used dust optical properties that overestimate dust absorption of shortwave radiation according to recent insights (1, 13). Results from these studies are thus not used here.) The dust radiative forcings at the surface and at top of atmosphere (TOA) are then given by

$$F_{surf} = M_d \sum_i F_{surf}^i \int_{D_{min}^i}^{D_{max}^i} \frac{dV_d}{dD_d} dD_d ; \quad (7a)$$

$$F_{TOA} = M_d \sum_i F_{TOA}^i \int_{D_{min}^i}^{D_{max}^i} \frac{dV_d}{dD_d} dD_d , \quad (7b)$$

where $M_d$ is the global emission rate of dust aerosols, and $i$ sums over the 7 particle bins used in ref. (19), for which $F_{surf}^i$ and $F_{TOA}^i$ are the mass-normalized radiative forcings in W m$^{-2}$ Tg$^{-1}$ Year at the surface and at TOA, and $D_{min}^i$ and $D_{max}^i$ are the bin's lower and upper size limits (SI Fig. 2).

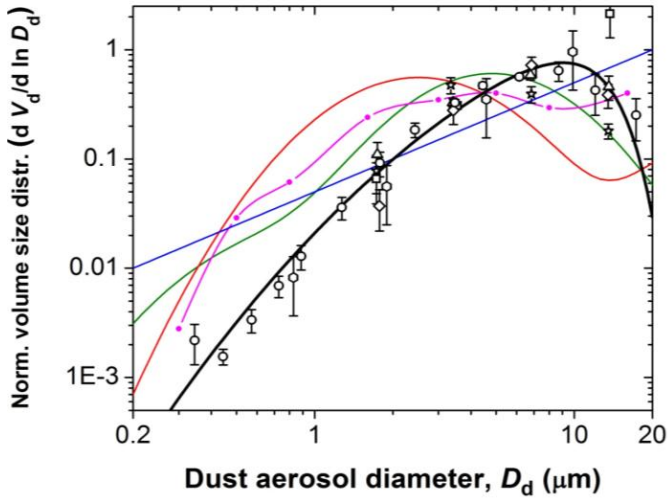

Fig. 3. The normalized volume size distribution of emitted dust aerosols used in 4 GCM studies (magenta line and circles (3), and blue (20), green (12), and red (13) lines). The thick black line denotes the theoretical PSD of Eq. 6, and symbols and error bars denote measurements as defined in Fig. 2.

The dust radiative forcing calculated with Eq. 7 is shown in Fig. 4 as a function of the global dust emission rate. For a given emission rate, the theoretical PSD of Eq. 6 predicts a surface radiative forcing that is a factor of ~2 – 6 less than predicted with empirical PSDs (3, 12, 13, 20), and also substantially smaller than predicted by GCM studies (13, 19, 51), with the exception of ref. (7). At TOA, the theoretical PSD produces a radiative forcing that is a factor of ~2 – 15 smaller than predicted with empirical PSDs and recent GCM studies (7, 13, 19, 51).

Fig. 4 thus indicates that the overestimation of the clay fraction causes GCMs either to underestimate the global dust emission rate or to overestimate the dust radiative cooling at the surface and especially at TOA. However, most GCM dust emission schemes are tuned to best match observations of the radiative characteristics of dust through comparisons against TOA radiative fluxes and dust aerosol optical depth (49, 52). Since these measurements are dominated by the long-lived and radiatively efficient clay fraction, correcting the overestimation of the clay fraction will likely increase the total mass of emitted dust aerosols at which GCMs produce optimal agreement with measurements (see SI Fig. 3 and Fig. 10 in ref. (49)). It is thus more plausible that GCMs substantially underestimate the global dust emission rate than that GCMs substantially overestimate dust radiative forcing. This result implies that the deposition flux of dust to ocean ecosystems may be substantially larger than previously thought, especially close to source regions.

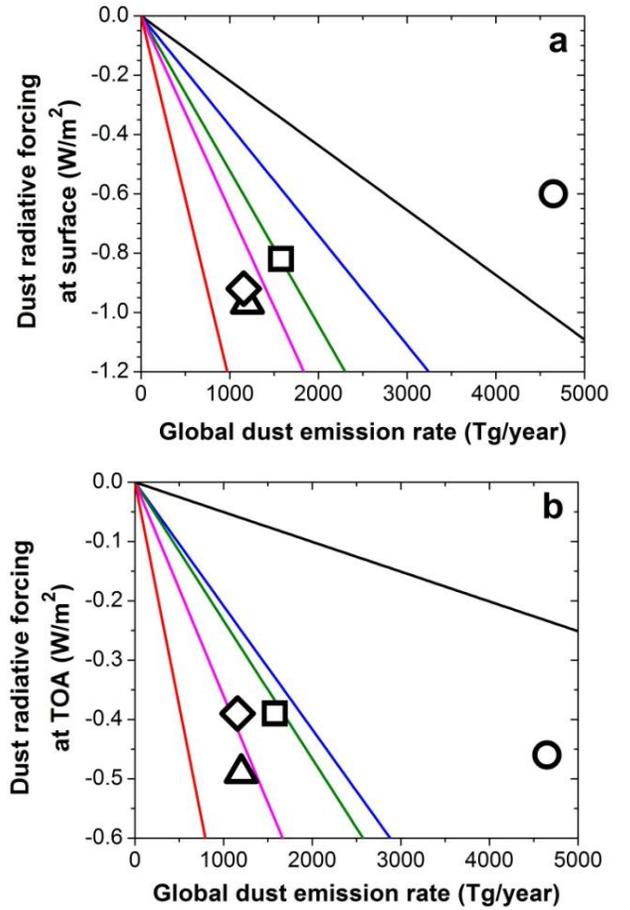

Fig. 4. The dust radiative forcing at the surface (**a**) and at TOA (**b**) calculated from Eq. (7) with the theoretical PSD (Eq. 6, solid black line) and with 4 empirical PSDs used in GCMs and plotted in Fig. 3 (magenta (3), blue (20), green (12), and red (13) lines). Included are radiative forcings calculated by 4 recent GCM studies (squares (19), circles (7), triangles (51), and diamonds (13)) that used dust shortwave absorption properties consistent with recent findings (1, 13).

Note that possible cancellation of errors leaves open the possibility that GCMs do get both the global dust emission rate and the dust radiative forcing approximately correct. For example, if GCMs substantially overestimate the deposition of clay aerosols, then the overestimation of the emitted clay fraction could result in roughly correct concentrations of atmospheric clay aerosols, thereby producing reasonable predictions of the global dust emission rate and dust radiative forcing.

Dust aerosols also affect Earth's radiative budget by serving as cloud nuclei (5) and thereby modifying cloud properties, a process known as the "indirect effect" (1). A GCM with detailed cloud physics would be required to assess the effect of the revised dust size distribution on estimates of the dust indirect effect.

**Summary and conclusions**

The present study indicates that dust emission is a scale invariant process (inset of Fig. 2), and uses this observation to derive a simple theoretical expression of the size distribution of emitted dust aerosols (Eqs. 5, 6). The theory is in excellent agreement with measurements (Fig. 2), and, when implemented in regional and global climate models, can resolve the substantial overestimation of the emitted clay fraction by these models (Fig. 3).

On local and regional scales, the overestimation of the emitted clay fraction by models has likely caused errors in the magnitude and, depending on local variables such as the suface albedo (4), potentially the sign of the modeled dust radiative forcing (50), which has implications for numerical weather forecasting and regional climate predictions in dusty regions. On a global scale, the overestimation of the emitted clay fraction has likely caused GCMs to underestimate the size of the global dust cycle (Fig. 4 and S3). This latter result implies that the deposition flux of dust to oceans, and the resulting effect on atmospheric greenhouse gas concentrations through the fertilization of marine biota (6), may be substantially larger than previously thought, especially close to dust source regions.

The theoretical model presented here could be applied to fragmentation in analogous physical systems where the creation of small fragments is limited by the presence of indivisible particles. This includes dust emission on Mars and the fragmentation of small asteroids (53), granular rocks (29), and other brittle materials with a granular or crystal structure.

**Acknowledgements**
I thank Annick Pouquet for pointing out the relevance of the scale invariance literature to dust emission, Stephanie Woodward and Yves Balkanski for providing me with the dust emission rates of the Woodward (2001) and Balkanski et al. (2007) studies, and Ted Zobeck, Marco Bittelli, Inez Fung, Natalie Mahowald, Charlie Zender, and an anonymous reviewer for many excellent suggestions. Comments by Shanna Shaked, John Seinfeld, Alexandra Jahn, Julie Theriault, Malcolm Brooks, Annick Pouquet, Charlie Zender, and Natalie Mahowald improved the clarity of the manuscript. The National Center for Atmospheric Research (NCAR) is sponsored by the National Science Foundation.

# Supporting Information

## Tables

Table S1. Summary of published measurements of the size-resolved vertical dust flux produced by eroding soils. Most studies reported the friction speed $u*$ (defined as the square root of the ratio of the wind stress and the air density), although Gillette et al. (1972, ref. (1)) only reported the wind speed $U$ at a height of 1.5 meters.

| Reference | Region | Soil Type | Wind speed range | Measured diameter range (μm) | Number of measurements | Used instrumentation |
|---|---|---|---|---|---|---|
| Gillette et al. (1972, ref. (1)) | Nebraska (US) | Sandy loam | $U$ = 4.2 m/s | 0.6 – 12 | 1 (sample 1) | 2 single-stage jet impactors at 1.5 and 6 m; size distribution was obtained from microscopy |
| Gillette (1974, ref. (2)), soil 1 | Texas (US) | Fine sand | $u*$ = 0.18 – 0.58 m/s | ~1.7 – 14 | 12 (runs 2 -13) | 2 single-stage jet impactors at 1.5 and 6 m; size distribution was obtained from microscopy |
| Gillette (1974, ref. (2)), soil 2 | Texas (US) | Fine sand | $u*$ = 0.49 – 0.78 m/s | ~1.7 – 14 | 4 (runs 1-4) | 2 single-stage jet impactors at 1.5 and 6 m; size distribution was obtained from microscopy |
| Gillette (1974, ref. (2)), soil 3 | Texas (US) | Loamy fine sand | $u*$ = 0.28 – 0.48 m/s | ~1.7 – 14 | 4 (runs 1-4) | 2 single-stage jet impactors at 1.5 and 6 m; size distribution was obtained from microscopy |
| Gillette et al. (1974, ref. (3)) | Texas (US) | Loamy fine sand | $u*$ = 0.24 – 0.78 m/s | ~1.7 – 28 | 8 (March 15, 27, 28 (2 data sets); April 2 (3 data sets), 18 | 2 single-stage jet impactors at 1.5 and 6 m; size distribution was obtained from microscopy |
| Sow et al. (2009, ref. (4)) | Niger | Not determined | $u*$ = 0.30 – 0.80 m/s | 0.3 – 20 | 3 (ME1, ME4, CE4) | 2 GRIMM optical particle sizers at 2.1 and 6.5 m |

Table S2: Clay fraction of atmospheric dust aerosols from measurements and GCM studies. The clay fraction is defined as the mass fraction of particles with diameters smaller than 2 μm in the PM20 size range.

| Reference | Study type | Geographical area | Clay fraction |
|---|---|---|---|
| Dubovik et al. (2002, ref. (5)) | Measurement (optical inversion) | Bahrain | 18 % |
| Dubovik et al. (2002, ref. (5)) | Measurement (optical inversion) | Solar Village (Saudi Arabia) | 13 % |
| Dubovik et al. (2002, ref. (5)) | Measurement (optical inversion) | Cape Verde | 17 % |
| Maring et al. (2003, ref. (6)) | Measurement (aerodynamic sampler) | Tenerife, Canary Islands | 21 % |
| Maring et al. (2003, ref. (6)) | Measurement (aerodynamic sampler) | Puerto Rico | 21 % |
| Reid et al. (2003, ref. (7)) | Measurement (literature average with aerodynamic samplers) | Various | 14 % |
| Reid et al. (2003, ref. (7)) | Measurement (literature average with optical inversion methods) | Various | 10 % |
| Tegen and Lacis (1996, ref. (8)) | GCM | Global | 48 % |
| Woodward (2001, ref. (9)) | GCM | Global | 26 % |
| Ginoux et al. (2001, ref. (10)) | GCM | Global | 39 % |
| Zender et al. (2003, ref. (11)) | GCM | Global | 37 % |
| Miller et al. (2004, ref. (12)) | GCM | Global | 63 % |
| Miller et al. (2006, ref. (13)) | GCM | Global | 20 % |
| Yue et al. (2010, ref. (14)) | GCM | Global | 20 % |

# Figures

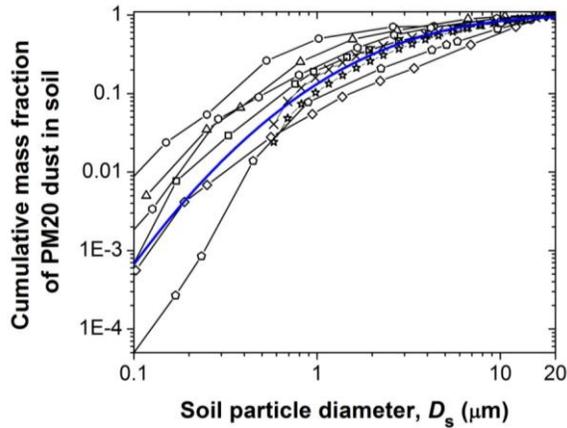 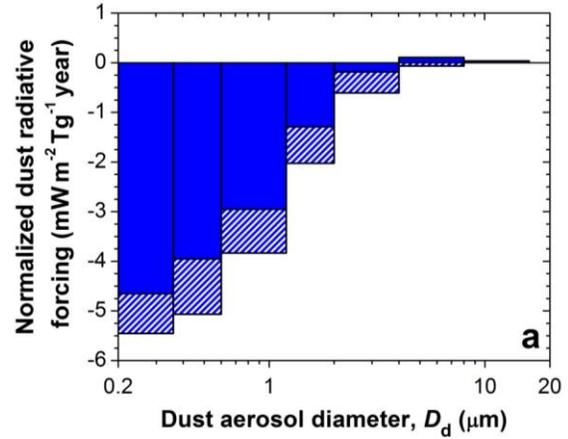

Fig. S1. Cumulative mass fraction of PM20 dust in two soils from Utah (stars and crosses respectively denote soils 00-U36 and 00-U37 in ref. (15)) and six soils from across the Sahara (squares, circles, triangles, diamonds, hexagons, and pentagons respectively denote soils 1 – 6 in ref. (16)). The thick blue line denotes the lognormal distribution (Eq. 4) with the log-normal parameters $\overline{D_s}$ and $\sigma_s$ taken as the average of the eight different soils from Table 1.

Fig. S2. The mass-normalized dust radiative forcing as a function of dust aerosol diameter at the surface (striped bars) and at TOA (solid bars). The mass-normalized radiative forcing was obtained by dividing the radiative forcing of individual particle size bins reported by ref. (13) by the corresponding yearly emitted aerosol mass.

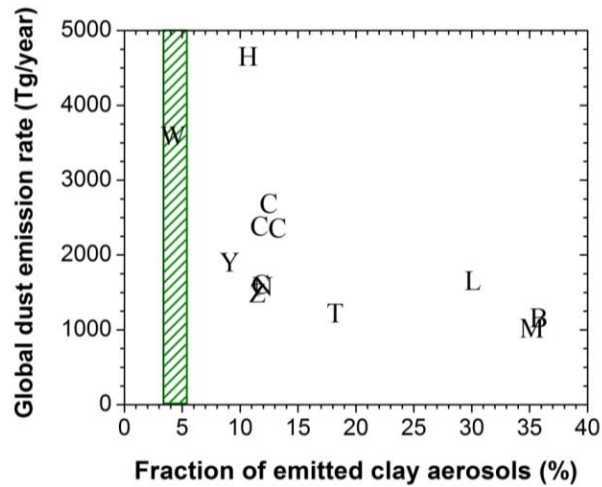

Fig. S3. Scatter plot of the global emission rate of PM20 dust aerosols versus the mass fraction of clay aerosols simulated by a range of GCMs. "B" denotes Balkanski et al. (17), "C" denotes the emitted clay fraction that gives optimal agreement with measurements for the 4 different preferred source descriptions used in Fig. 10 of Cakmur et al. (18), "H" denotes Mahowald et al. (19), "L" denotes Luo et al. (20), "M" denotes Miller et al. (2004, ref. (12)), "N" denotes Miller et al. (2006, ref. (13)), "T" denotes Tegen and Lacis (8), "W" denotes Woodward (9), "Y" denotes Yue et al. (14), and "Z" denotes Zender et al. (11). Note that the largest aerosols simulated by some of these studies are smaller than 20 μm (12, 13, 18-20), and that the emission rate of Woodward (9) includes particles which are immediately deposited back to the surface without any interaction with the model. The shaded green bar represents the emitted clay fraction of 4.4 ± 1.0 % predicted from Eq. (6) with the average values and standard errors of $\overline{D_s}$, $\sigma_s$, and $\lambda$ listed in Table 1.

**Supplementary references**